\documentclass[preprint,12pt]{elsarticle}
\usepackage{graphicx}
\usepackage{amssymb}
\usepackage{amsmath}
\usepackage{subfig}
\usepackage{multirow}
\usepackage{colortbl,color}
\usepackage{algorithm}
\usepackage{amsthm}
\usepackage{hyperref}
\biboptions{comma,sort&compress}
\journal{ }

\begin{document}

\begin{frontmatter}
\title{DSBGK Method to Incorporate the CLL Reflection Model and to Simulate Gas Mixtures}
\author[KFUPM]{Jun~Li\footnote[1]{e-mail: lijun04@gmail.com \\ To view more DSBGK results, please visit \href{http://modelnanogasflows.com/}{\textit{NanoGasSim}}.}}
\address[KFUPM]{Center for Integrative Petroleum Research, \\ College of Petroleum Engineering and Geosciences, \\ King Fahd University of Petroleum $\&$ Minerals, Saudi Arabia}
\begin{abstract}
Molecular reflections on usual wall surfaces can be statistically described by the Maxwell diffuse reflection model, which has been successfully applied in the DSBGK simulations. We develop the DSBGK algorithm to implement the Cercignani-Lampis-Lord (CLL) reflection model, which is widely applied to polished surfaces and used particularly in modeling space shuttles to predict the heat and force loads exerted by the high-speed flows around the surfaces. We also extend the DSBGK method to simulate gas mixtures and high contrast of number densities of different components can be handled at a cost of memory usage much lower than that needed by the DSMC simulations because the average numbers of simulated molecules of different components per cell can be equal in the DSBGK simulations.   
\end{abstract}
\end{frontmatter}
\section{Introduction}\label{s:intro}
In the molecular reflection processes on wall surfaces, the CLL reflection model \cite{CL1971}-\cite{CLL1991} based on a probability distribution is usually employed to model the molecular reflection behavior when the reflecting molecular velocity $\vec c_{\rm r}$ is statistically correlated with the incoming velocity $\vec c_{\rm i}$. The complicated form of the CLL model makes its implementations difficult in numerical methods (e.g., molecular simulation methods or direct discretization methods of the Boltzmann equation) that require the value of distribution function $f$ while the application of CLL model in the DSMC method \cite{Bird1994} is convenient \cite{CLL1991} because the value of $f$ is not needed. The DSBGK method \cite{Li2011RGD}-\cite{Li2012arXiv} is a molecular simulation method and meanwhile requires the value of $f$. As discussed in \cite{Li2012arXiv}, the distribution function $f_{S^{'}}(\vec c_{\rm i})$ of incoming molecules with velocity $\vec c_{\rm i}$ in a local Cartesian reference frame $S^{'}$ that moves together with the wall boundary at $\vec u_{\rm wall}$ is required to update $f$ after each molecular reflection using double integral and the integration is analytically impossible even if $f_{S^{'}}(\vec c_{\rm i})$ takes the simple form of Maxwell distribution function. Nevertheless, we can make full use of the features of molecular simulation method and use the transient distribution of $f_{S^{'}}(\vec c_{\rm i})$, namely a summation of Dirac delta functions, to complete the integration with respect to $\vec c_{\rm i}$. Statistically, the transient distribution is valid according to the analysis of consistency between the DSMC method and Boltzmann equation \cite{Li2012arXiv}. Using the transient \textit{discrete} distribution of incoming molecules to update the value of a \textit{continuous} distribution of reflecting molecules might lead to numerical instability but a similar idea has been successfully applied to compute the incoming number flux rate $N_{\rm in}$ to implement the Maxwell diffuse reflection model by using a large number of simulated molecules per cell to avoid instability \cite{Li2014arXiv}. Here, we present an algorithm to implement the CLL reflection model with theoretical analysis.  

Compared to the DSMC method, the observed remarkable advantage of the DSBGK method is the high efficiency in low-speed (low Mach number in general) cases as shown in the benchmark studies \cite{Li2012arXiv}-\cite{Li2014arXiv}, the permeability study of shale gas as a function of pore pressure \cite{LiSultan2015JNGSE}-\cite{LiSultan2015IPTC}, and the study of thermal transpiration flows with validations by experimental data for several gas species over a wide range of Knudsen number \cite{LiCai2017arXiv}. Another potential advantage of the DSBGK method is the capability to simulate gas mixtures with high contrast of number densities of different components at much lower memory usage (consequently with much higher efficiency even if the Mach number is not low) compared to that needed by the DSMC simulations. For example, to simulate a mixture of gas $\sigma_1$ and $\sigma_2$ with a number density ratio $n_{\sigma_1}/n_{\sigma_2}=100$, DSMC simulations usually employ about 20 simulated molecules of component $\sigma_2$ per cell and then needs about $2000$ simulated molecules of component $\sigma_1$ per cell (2020 per cell in total), which implies a very high memory usage. By contrast, DSBGK simulations can use about 20 simulated molecules per cell for both components $\sigma_1$ and $\sigma_2$ (40 per cell in total) because the numbers of real molecules represented by each simulated molecule for different components can be arbitrarily specified instead of \textit{must} being equal for all components as required in the DSMC simulations. 

\section{DSBGK Method}\label{s:DSBGK method}
We consider the gas flows of single component in the absence of external body force. The BGK equation \cite{BGK1954} can be written as a Lagrangian form:
\begin{equation}\label{eq:BGK}
    \dfrac{{\rm d}f}{{\rm d}t}=\dfrac{\partial f}{\partial t}+\vec c\cdot\dfrac{\partial f}{\partial \vec x}=\upsilon(f^{\rm eq}-f),
\end{equation}
where $f(t, \vec x, \vec c)$ is the unknown probability distribution function, $t$ the time, $\vec x$ the spatial coordinate, $\vec c$ the molecular velocity and, the coefficient $\upsilon$ is appropriately selected to satisfy the coefficient of dynamic viscosity or heat conduction \cite{VincentiKruger1965} (detailed in \cite{Li2012arXiv}) and the Maxwell distribution function $f^{\rm eq}$ is:
\begin{equation}\label{eq:feq}
    f^{\rm eq}(n, \vec u, T)=n(\dfrac{m}{2\pi k_{\rm B}T})^{3/2}\exp[\dfrac{-m(\vec c-\vec u)^2}{2k_{\rm B}T}],
\end{equation}
where $f^{\rm eq}$ essentially is a function of $t$, $\vec x$ and $\vec c$ although notation $f^{\rm eq}(n, \vec u, T)$ is used for the convenience of discussion, $m$ is the molecular mass, $k_{\rm B}$ is the Boltzmann constant and, the number density $n$, flow velocity $\vec u$ and temperature $T$ are functions of $t$ and $\vec x$ and defined using the integrals of $f$ with respect to $\vec c$ as shown in Eq.~\eqref{eq:nuT-sigma_i}.

The DSBGK method is proposed in \cite{Li2011RGD} and detailed in \cite{Li2012arXiv}, where the extension to problems with external force is discussed. The simulation process is divided into a series of time steps $\Delta t$ and the computational domain is divided into many regular or irregular cells. The cell size $\Delta L_{\rm cell}$ and $\Delta t$ are selected the same as in the DSMC method. Each simulated molecule $l$ carries four molecular variables: position $\vec x_l$, molecular velocity $\vec c_l$,  number $N_l$ of real molecules represented by the simulated molecule $l$, and $F_l$ that is equal to $f(t, \vec x_l, \vec c_l)$. The variables $n_{{\rm tr,}k}, \vec u_{{\rm tr,}k}, T_{{\rm tr,}k}$ of each cell $k$ are updated using $\vec x_l, \vec c_l$ and the increment of $N_l$ in the cell $k$ based on the mass, momentum and energy conservation laws of the intermolecular collision process. These cell's variables are simultaneously used in turn to update the molecular variables based on the BGK equation and an extrapolation \cite{Li2012arXiv} of the acceptance-rejection scheme. The DSBGK method is a molecular simulation method and theoretically all macroscopic quantities (e.g., cell's variables) of interest should be computed using the transient molecular variables as in the DSMC method. Nevertheless, the transitional cell's variables $n_{{\rm tr,}k}, \vec u_{{\rm tr,}k}, T_{{\rm tr,}k}$ are introduced in the DSBGK method and used in place of the original $n_k, \vec u_k, T_k$, which are defined by the transient molecular variables inside the cell $k$, to reduce stochastic noise. $n_{{\rm tr,}k}, \vec u_{{\rm tr,}k}, T_{{\rm tr,}k}$ can evolve smoothly and will converge to $n_k, \vec u_k, T_k$, respectively, as discussed after Eq.~(13) of \cite{Li2012arXiv}.

\subsection{Summary of the DSBGK algorithm}\label{ss:algorithm summary}
1. At the initial state, many cells and simulated molecules are generated and then, initial values are assigned to cell's variables $n_{{\rm tr,}k}, \vec u_{{\rm tr,}k}, T_{{\rm tr,}k}$ and molecular variables $\vec x_l, \vec c_l, F_l, N_l$, respectively, according to the initial distribution $f_0=f^{\rm eq}(n_0, \vec u_0,T_0)$.

2. Each simulated molecule moves uniformly and in a straight line before randomly reflecting at the wall surfaces according to a specified reflection model (e.g., Maxwell diffuse reflection model or CLL model). During each $\Delta t$, the trajectory of each particular molecule $l$ may be divided into several segments by the cell's interfaces. Then, $\vec x_l, F_l, N_l$ are \textit{deterministically} updated along each segment in sequence at the moving direction. When encountering wall boundaries, $\vec c_l$ is randomly updated according to the reflection model and then $F_l$ is updated correspondingly. Simulated molecules are removed from the computational domain when moving across the open boundaries during each $\Delta t$ and then new simulated molecules are generated after each $\Delta t$ at the open boundaries according to the specified pressures. The variables $n_{{\rm tr,}k}, \vec u_{{\rm tr,}k}, T_{{\rm tr,}k}$ of each cell $k$ are updated after each $\Delta t$ according to the conservation laws.

3. After convergence, $n_{{\rm tr,}k}, \vec u_{{\rm tr,}k}, T_{{\rm tr,}k}$ are used as the discrete solutions of the BGK equation at steady state.

\section{An Algorithm for the CLL Reflection Model at Boundary}\label{s:CLL model}
In the reflection models of wall boundary, $\vec c_l$ and then $F_l$ are changed after molecular reflection at $\vec x_l$ on the wall. $N_l$ remains unchanged to conserve the mass. We use the subscripts 2 and 3 for the tangential directions and 1 for the outer-normal direction of the wall surface in $S^{'}$ and use $x,y,z$ as subscripts in the global Cartesian reference frame $S$ when needed. The subscript $l$ is omitted in the notations of the incoming velocity $\vec c_{\rm i}$ and reflecting velocity $\vec c_{\rm r}$, which are observed in $S^{'}$. $\vec c_{\rm r}$ is randomly generated the same as in the DSMC simulations and then $\vec c_l$ is updated to $\vec c_{\rm r}+\vec u_{\rm wall}$ (see the details in \cite{Li2012arXiv}). 

As discussed in \cite{Li2012arXiv}, $F_l$ is updated to $F^{\rm new}_l=f(t, \vec x_l, \vec c^{\rm new}_l)=f(t, \vec x_l, \vec c_{\rm r}+\vec u_{\rm wall})$ after getting $\vec c_{\rm r}$. Note that $F_l$ is the representative value of $f$, which is different from the scatter kernel $R$ that is used to generate $\vec c_{\rm r}$ for each particular reflection process. Generally speaking, $f$ is related to the mass flux rate but $R$ has nothing to do with the mass flux rate. Usually, $R$ describes the probability distribution of $\vec c_{\rm r}$ inside the half velocity space ($c_{{\rm r},1}=\vec c_{\rm r}\cdot\vec e_{\rm n}>0$, where $\vec e_{\rm n}$ is the outer-normal unit vector of the wall) as a function that generally depends on the wall temperature $T_{\rm wall}$, accommodation coefficients $\alpha_{\rm n}, \alpha_\tau$ and the incoming velocity $\vec c_{\rm i}$. So, we have $R=R(\vec c_{\rm i}{\to}\vec c_{\rm r})$ that contains $T_{\rm wall}, \alpha_{\rm n}, \alpha_\tau$ as coefficients. $R$ satisfies the normalization condition $\int_{\vec c_{\rm r}\cdot\vec e_{\rm n}>0}R(\vec c_{\rm i}{\to}\vec c_{\rm r}){\rm d}\vec c_{\rm r}=1$ for arbitrary $\vec c_{\rm i}$, where $R(\vec c_{\rm i}{\to}\vec c_{\rm r}){\rm d}\vec c_{\rm r}$ is the probability for the molecule coming at $\vec c_{\rm i}$ to reflect into the velocity space element ${\rm d}\vec c_{\rm r}$ around $\vec c_{\rm r}$. The transformation between $f$ at the wall location and $R$ is discussed below.

We introduce $f_{S^{'}}(\vec c)$ as the equivalent distribution function of $f$ observed in $S^{'}$ at the reflection point $\vec x_l$ and at the current moment $t$, which means $f_{S^{'}}(\vec c)=f(t, \vec x_l, \vec c+\vec u_{\rm wall})$. After getting the formula of $f_{S^{'}}(\vec c)$, we have $F^{\rm new}_l=f_{S^{'}}(\vec c_{\rm r})$. $f_{S^{'}}(\vec c_{\rm i})|_{\vec c_{\rm i}\cdot\vec e_{\rm n}<0}$ is the distribution of incoming molecules in the cell adjacent to the reflection position. $f_{S^{'}}(\vec c_{\rm r})|_{\vec c_{\rm r}\cdot\vec e_{\rm n}>0}$ is the distribution of reflecting molecules and related to $R$ as introduced in \cite{Shen2005}:
\begin{equation}\label{eq:kernel}
    f_{S^{'}}(\vec c_{\rm r})(\vec c_{\rm r}\cdot\vec e_{\rm n}){\rm d}\vec c_{\rm r}=-\int_{\vec c_{\rm i}\cdot\vec e_{\rm n}<0}R(\vec c_{\rm i}{\to}\vec c_{\rm r})f_{S^{'}}(\vec c_{\rm i})(\vec c_{\rm i}\cdot\vec e_{\rm n}){\rm d}\vec c_{\rm i}{\rm d}\vec c_{\rm r}.
\end{equation}
Taking integration of Eq.~\eqref{eq:kernel} with respect to $\vec c_{\rm r}$ over its half velocity space and using the normalization condition of $R(\vec c_{\rm i}{\to}\vec c_{\rm r})$, we get:
\begin{equation}\label{eq:massbalance}
\begin{aligned}
    &\int_{\vec c_{\rm r}\cdot\vec e_{\rm n}>0}f_{S^{'}}(\vec c_{\rm r})(\vec c_{\rm r}\cdot\vec e_{\rm n}){\rm d}\vec c_{\rm r} \\
    &=-\int_{\vec c_{\rm r}\cdot\vec e_{\rm n}>0}\int_{\vec c_{\rm i}\cdot\vec e_{\rm n}<0}R(\vec c_{\rm i}{\to}\vec c_{\rm r})f_{S^{'}}(\vec c_{\rm i})(\vec c_{\rm i}\cdot\vec e_{\rm n}){\rm d}\vec c_{\rm i}{\rm d}\vec c_{\rm r} \\
    &=-\int_{\vec c_{\rm i}\cdot\vec e_{\rm n}<0}f_{S^{'}}(\vec c_{\rm i})(\vec c_{\rm i}\cdot\vec e_{\rm n}){\rm d}\vec c_{\rm i},
\end{aligned}
\end{equation}
which represents the mass conservation of molecular reflection process.

In the CL reflection model \cite{CL1971}, the scatter kernel $R$ is the product of three independent parts respectively related to the three velocity components:
\begin{equation}\label{eq:kernelCL}
\begin{aligned}
    R_{\rm CL}(\vec c_{\rm i}{\to}\vec c_{\rm r})=
    &\dfrac{1}{\sqrt{\pi\alpha_{\tau}}}\exp[\dfrac{-(\tilde{c}_{{\rm r,}2}-\sqrt{1-\alpha_\tau}\tilde{c}_{{\rm i,}2})^2}{\alpha_\tau}]\times
    \\ &\dfrac{1}{\sqrt{\pi\alpha_{\tau}}}\exp[\dfrac{-(\tilde{c}_{{\rm r,}3}-\sqrt{1-\alpha_\tau}\tilde{c}_{{\rm i,}3})^2}{\alpha_\tau}]\times
    \\ &\dfrac{\tilde{c}_{{\rm r,}1}}{\pi\alpha_{\rm n}}\exp[\dfrac{-(\tilde{c}_{{\rm r,}1}^2+(1-\alpha_{\rm n})
    \tilde{c}_{{\rm i,}1}^2)}{\alpha_{\rm n}}]\times
    \\ &\int_{0}^{2\pi}\exp[\dfrac{2\sqrt{1-\alpha_{\rm n}}\tilde{c}_{{\rm r,}1}|\tilde{c}_{{\rm i,}1}|}{\alpha_{\rm n}}
    \cos\theta]{\rm d}\theta,
\end{aligned}
\end{equation}
where $|\tilde{c}_{{\rm i,}1}|$ is the absolute value of the normalized incoming component $\dfrac{c_{{\rm i,}1}}{\sqrt{2k_{\rm B}T_{\rm wall}/m}}$ with $c_{{\rm i,}1}<0$. The generating algorithm of $\vec c_{\rm r}$ was proposed in \cite{CLL1991} based on Eq.~\eqref{eq:kernelCL} and is referred to as the CLL reflection model. Small modification was proposed in \cite{Li2012arXiv} to improve the efficiency of implementing the CLL algorithm.  

As discussed in the analysis of consistency between the DSMC method and Boltzmann equation \cite{Li2012arXiv}, we assume that the differences between the coordinates $\vec x_l$ of those simulated molecules located inside the cell $k$ and the reflection positions around the cell $k$ are negligible. Then, for each molecular reflection around the cell $k$, the transient $f_{S^{'}}(\vec c_{\rm i})$ of the incoming molecules is a summation of Dirac delta functions with $\delta(\vec 0){\rm d}\vec c_i=1$ as follows: 
\begin{equation}\label{eq:fBci}
\begin{aligned}
    f_{S^{'}}(\vec c_{\rm i})=\sum_{l\in{\rm cell}k}\delta(\vec c_l-\vec u_{\rm wall}-\vec c_{\rm i})N_l/\Delta V_k,
\end{aligned}
\end{equation}
where $\Delta V_k$ is the volume of cell $k$ and $\sum_{l\in{\rm cell}k}$ is the summation over all simulated molecules located inside the cell $k$. To make the algorithm general and robust, we use the same set $\left\{\vec c_l, N_l\left| l\in{\rm cell}k\right.\right\}$, which is stored and updated at the \textit{beginning} of each $\Delta t$ for each cell $k$ adjacent to wall, to compute the same transient $f_{S^{'}}(\vec c_{\rm i})$ for all subsequent molecular reflections around the cell $k$ during the concerned $\Delta t$ because the dynamic set becomes not representative when simulated molecules are updated in an artificially specified order particularly in simulating open problems, where new simulated molecules are generated at the end of each time step, before which the dynamic set close to open boundary is not complete. Substituting Eq.~\eqref{eq:fBci} into Eq.~\eqref{eq:kernel}, we get: 
\begin{equation}\label{eq:fBcr}
\begin{aligned}
    f_{S^{'}{\rm ,CL}}(\vec c_{\rm r})=\sum_{l\in{\rm cell}k \atop (\vec c_l-\vec u_{\rm wall})\cdot\vec e_{\rm n}<0}R_{\rm CL}((\vec c_l-\vec u_{\rm wall}){\to}\vec c_{\rm r})[(\vec c_l-\vec u_{\rm wall})\cdot\vec e_{\rm n}]\dfrac{-N_l}{\Delta V_kc_{{\rm r},1}}.
\end{aligned}
\end{equation}
For each molecular reflection around cell $k$, the reflecting velocity $\vec c_{\rm r}=(c_{{\rm r},1},c_{{\rm r},2},c_{{\rm r},3})$ will be generated according to the CLL model and then we update $F_l$ to $f_{S^{'}{\rm ,CL}}(\vec c_{\rm r})$ computed using Eq.~\eqref{eq:fBcr}, where $R_{\rm CL}$ needs to be calculated for each term $l$ of the summation by numerical integration with respect to $\theta$ using Eq.~\eqref{eq:kernelCL}.  

The applications of the specular reflection model and the Maxwell diffuse reflection model are discussed in \cite{Li2012arXiv}-\cite{Li2014arXiv}.

\section{Extension to Gas Mixtures}\label{s:gas mixtures}
\subsection{Governing equation}\label{ss:multiBGK equation}
We extend the DSBGK method for simulating gas mixtures without chemical reaction based on a consistent BGK-type model \cite{AndriesAokiPerthame2002}, which satisfies several fundamental properties. This extension involves very few modifications to the original DSBGK algorithm and other extensions are possible by using different BGK-type equations . 

As in the original BGK equation, the macroscopic quantities of each component $\sigma_i\in[\sigma_1, \sigma_N]$ (\textit{note}: subscript $N$ is used for the total number of components) are defined using the distribution function $f_{\sigma_i}(t, \vec x, \vec c)$: 
\begin{equation}\label{eq:nuT-sigma_i}
    \begin{cases}
    n_{\sigma_i}=\int_{\mathbb{R}^3}f_{\sigma_i}{\rm d}\vec c \\
    \vec u_{\sigma_i}=\dfrac{1}{n_{\sigma_i}}\int_{\mathbb{R}^3}\vec cf_{\sigma_i}{\rm d}\vec c \\
    T_{\sigma_i}=\dfrac{2\epsilon_{\sigma_i}}{3k_{\rm B}}=\dfrac{m_{\sigma_i}}{3k_{\rm B}n_{\sigma_i}}\int_{\mathbb{R}^3}(\vec c-\vec u_{\sigma_i})^2f_{\sigma_i}{\rm d}\vec c,
    \end{cases}
\end{equation}
where $\epsilon_{\sigma_i}$ is the internal energy per molecule of component $\sigma_i$ associated with random thermal motions. Total number density $n$, mean flow velocity $\vec u$ and temperature $T$ of the mixture can be defined using $n_{\sigma_i}$, $\vec u_{\sigma_i}$, $T_{\sigma_i}$ and molecular mass $m_{\sigma_i}$ of all components. The evolution of $f_{\sigma_i}$ is as follows \cite{AndriesAokiPerthame2002}:   
\begin{equation}\label{eq:multiBGK}
    \dfrac{{\rm d}f_{\sigma_i}}{{\rm d}t}=\dfrac{\partial f_{\sigma_i}}{\partial t}+\vec c\cdot\dfrac{\partial f_{\sigma_i}}{\partial \vec x}=\upsilon_{\sigma_i}(f^{\rm eq}_{\sigma_i}-f_{\sigma_i}),
\end{equation}
where the total collision frequency is $\upsilon_{\sigma_i}=\sum_{\sigma_j=\sigma_1}^{\sigma_N}\upsilon_{\sigma_i\sigma_j}n_{\sigma_j}$ and 
\begin{equation}\label{eq:multifeq}
    f^{\rm eq}_{\sigma_i}(n_{\sigma_i},\vec u^{\rm eq}_{\sigma_i},T^{\rm eq}_{\sigma_i})=n_{\sigma_i}(\dfrac{m_{\sigma_i}}{2\pi k_{\rm B}T^{\rm eq}_{\sigma_i}})^{3/2}\exp[\dfrac{-m_{\sigma_i}(\vec c-\vec u^{\rm eq}_{\sigma_i})^2}{2k_{\rm B}T^{\rm eq}_{\sigma_i}}]
\end{equation}
and the auxiliary quantities $\vec u^{\rm eq}_{\sigma_i}$, $\epsilon^{\rm eq}_{\sigma_i}=\dfrac{3k_{\rm B}T^{\rm eq}_{\sigma_i}}{2}$ are
\begin{equation}\label{eq:multiueq}
    m_{\sigma_i}\upsilon_{\sigma_i}\vec u^{\rm eq}_{\sigma_i}=m_{\sigma_i}\upsilon_{\sigma_i}\vec u_{\sigma_i}+\sum_{\sigma_j=\sigma_1}^{\sigma_N}2\mu_{\sigma_i\sigma_j}\chi_{\sigma_i\sigma_j}n_{\sigma_j}(\vec u_{\sigma_j}-\vec u_{\sigma_i}) 
\end{equation}
and 
\begin{equation}\label{eq:multiepsiloneq}
\begin{aligned}
    \upsilon_{\sigma_i}\epsilon^{\rm eq}_{\sigma_i}= & \upsilon_{\sigma_i}\epsilon_{\sigma_i}-\dfrac{m_{\sigma_i}\upsilon_{\sigma_i}}{2}(\vec u^{\rm eq}_{\sigma_i}-\vec u_{\sigma_i})^2 \\ 
    & +\sum_{\sigma_j=\sigma_1}^{\sigma_N}\dfrac{4\mu_{\sigma_i\sigma_j}\chi_{\sigma_i\sigma_j}n_{\sigma_j}}{m_{\sigma_i}+m_{\sigma_j}}[\epsilon_{\sigma_j}-\epsilon_{\sigma_i}+\dfrac{m_{\sigma_j}(\vec u_{\sigma_j}-\vec u_{\sigma_i})^2}{2}],
\end{aligned}
\end{equation}
where $\mu_{\sigma_i\sigma_j}=\dfrac{m_{\sigma_i}m_{\sigma_j}}{m_{\sigma_i}+m_{\sigma_j}}$ is the reduced mass and $\chi_{\sigma_i\sigma_j}$ is the interaction coefficient between components $\sigma_i$ and $\sigma_j$. The coefficients $\upsilon_{\sigma_i\sigma_j}$ and $\chi_{\sigma_i\sigma_j}$ are defined using the interaction potential \cite{AndriesAokiPerthame2002}. 

During each $\Delta t$, the mass increment $\Delta M_{k,\sigma_i}$ of component $\sigma_i$ in the cell $k$ due to intermolecular collisions with all components is: 
\begin{equation}\label{eq:massincrement-sigma_i}
\begin{aligned}
    \Delta M_{k,\sigma_i} & = \Delta t\Delta V_k\int_{\mathbb{R}^3}m_{\sigma_i}\upsilon_{\sigma_i}(f^{\rm eq}_{\sigma_i}-f_{\sigma_i}){\rm d}\vec c \\
                                        & = \Delta t\Delta V_km_{\sigma_i}\upsilon_{\sigma_i}(n_{\sigma_i}-n_{\sigma_i}) \\
                                        & \equiv 0,
\end{aligned}
\end{equation}
which is consistent with the mass conservation. 

During each $\Delta t$, the momentum increment $\Delta P_{k,\sigma_i}$ of component $\sigma_i$ in the cell $k$ due to intermolecular collisions with all components is: 
\begin{equation}\label{eq:momentumincrement-sigma_i}
\begin{aligned}
    \Delta P_{k,\sigma_i} & = \Delta t\Delta V_k\int_{\mathbb{R}^3}(m_{\sigma_i}\vec c)\upsilon_{\sigma_i}(f^{\rm eq}_{\sigma_i}-f_{\sigma_i}){\rm d}\vec c \\
                                        & = \Delta t\Delta V_k\upsilon_{\sigma_i}n_{\sigma_i}m_{\sigma_i}(\vec u^{\rm eq}_{\sigma_i}-\vec u_{\sigma_i}) \\
                                        & = \Delta t\Delta V_k\sum_{\sigma_j=\sigma_1}^{\sigma_N}2n_{\sigma_i}n_{\sigma_j}\mu_{\sigma_i\sigma_j}\chi_{\sigma_i\sigma_j}(\vec u_{\sigma_j}-\vec u_{\sigma_i}),
\end{aligned}
\end{equation}
where Eq.~\eqref{eq:multiueq} is substituted. $\Delta P_{k,\sigma_i}$ could be nonzero due to momentum exchange between components via intermolecular collisions but the global momentum conservation is satisfied as $\sum_{\sigma_i=\sigma_1}^{\sigma_N}\Delta P_{k,\sigma_i}\equiv 0$. 

During each $\Delta t$, the energy increment $\Delta E_{k,\sigma_i}$ of component $\sigma_i$ in the cell $k$ due to intermolecular collisions with all components is: 
\begin{equation}\label{eq:energyincrement-sigma_i}
\begin{aligned}
    \Delta E_{k,\sigma_i} = & \Delta t\Delta V_k\int_{\mathbb{R}^3}\dfrac{m_{\sigma_i}(\vec c)^2}{2}\upsilon_{\sigma_i}(f^{\rm eq}_{\sigma_i}-f_{\sigma_i}){\rm d}\vec c \\
                                        = & \Delta t\Delta V_k\upsilon_{\sigma_i}n_{\sigma_i}[\epsilon^{\rm eq}_{\sigma_i}+\dfrac{m_{\sigma_i}}{2}(\vec u^{\rm eq}_{\sigma_i})^2-\epsilon_{\sigma_i}-\dfrac{m_{\sigma_i}}{2}(\vec u_{\sigma_i})^2] \\
                                        = & \Delta t\Delta V_k\sum_{\sigma_j=\sigma_1}^{\sigma_N}\dfrac{2n_{\sigma_i}n_{\sigma_j}\mu_{\sigma_i\sigma_j}\chi_{\sigma_i\sigma_j}}{m_{\sigma_i}+m_{\sigma_j}} \\
                                           & [2\epsilon_{\sigma_j}-2\epsilon_{\sigma_i}+(\vec u_{\sigma_j}-\vec u_{\sigma_i})\cdot(m_{\sigma_i}\vec u_{\sigma_i}+m_{\sigma_j}\vec u_{\sigma_j})],
\end{aligned}
\end{equation}
where Eqs.~\eqref{eq:multiueq} and  \eqref{eq:multiepsiloneq} are substituted. $\Delta E_{k,\sigma_i}$ could be nonzero due to energy exchange between components via intermolecular collisions but the global energy conservation is satisfied as $\sum_{\sigma_i=\sigma_1}^{\sigma_N}\Delta E_{k,\sigma_i}\equiv 0$.

\subsection{DSBGK algorithm}\label{ss:multiDSBGK algorithm}
In the DSBGK simulations of gas mixtures, each molecule $l$ with a component index $\sigma_l\in[\sigma_1,\sigma_N]$ (\textit{note}: we use the notation $\sigma_l$ as a component index associated with the simulated molecule $l$ for simplicity but $\sigma_l$ of the first simulated molecule with $l=1$ is \textit{not} necessary equal to $\sigma_1$ as the first component, for example) has four variables: $\vec x_l$, $\vec c_l$, $N_l$ and $F_l=f_{\sigma_l}(t,\vec x_l,\vec c_l)$ as in the original algorithm. The magnitude of initial $N_l$ of component $\sigma_l=\sigma_i$ could be proportional to the initial number density $n_{\sigma_i,0}$ such that the average numbers of simulated molecules per cell are almost equal for all components. Each cell $k$ has three original variables $n_{{\rm tr},k,\sigma_i}$, $\vec u_{{\rm tr},k,\sigma_i}$, $T_{{\rm tr},k,\sigma_i}$ and two additional auxiliary variables $\vec u^{\rm eq}_{{\rm tr},k,\sigma_i}$, $T^{\rm eq}_{{\rm tr},k,\sigma_i}$ for each component $\sigma_i$. 

At the initial state with distributions of $n_{\sigma_i,0}$, $\vec u_{\sigma_i,0}$ and $T_{\sigma_i,0}$,  we have $n_{{\rm tr},k,\sigma_i}=n_{\sigma_i,0}$, $\vec u_{{\rm tr},k,\sigma_i}=\vec u^{\rm eq}_{{\rm tr},k,\sigma_i}=\vec u_{\sigma_i,0}$ and $T_{{\rm tr},k,\sigma_i}=T^{\rm eq}_{{\rm tr},k,\sigma_i}=T_{\sigma_i,0}$. The values of molecular variables of each component $\sigma_i$ are determined according to the initial distribution $f_{\sigma_i,0}=f^{\rm eq}_{\sigma_i}(n_{\sigma_i,0},\vec u_{\sigma_i,0}, T_{\sigma_i,0})$. 

During each $\Delta t$, molecular variables are updated using Eq.~\eqref{eq:multiBGK} with $f^{\rm eq}_{\sigma_i}(n_{{\rm tr},k,\sigma_i},\vec u^{\rm eq}_{{\rm tr},k,\sigma_i},T^{\rm eq}_{{\rm tr},k,\sigma_i})$ as in the original DSBGK algorithm . The error between the numerical mass increment and theoretical mass increment $\Delta M_{k,\sigma_i}$ of component $\sigma_i$ in the cell $k$ due to intermolecular collisions with all components is 
\begin{equation}\label{eq:massincrementerr-sigma_i}
\begin{aligned}
    \Delta M^{\rm err}_{k,\sigma_i}=m_{\sigma_i}\sum_{l\in{\rm cell}k \atop \sigma_l=\sigma_i}\Delta_kN_l-\Delta M_{k,\sigma_i}=m_{\sigma_i}\sum_{l\in{\rm cell}k \atop \sigma_l=\sigma_i}\Delta_kN_l,
\end{aligned}
\end{equation}
where $\Delta_kN_l$ is the number increment of real molecules of class $\vec c_l$ of component $\sigma_l$ due to intermolecular collisions with all components inside the cell $k$ during the current time step \cite{Li2012arXiv}. The error between the numerical momentum increment and theoretical momentum increment $\Delta P_{k,\sigma_i}$ of component $\sigma_i$ in the cell $k$ due to intermolecular collisions with all components is 
\begin{equation}\label{eq:momentumincrementerr-sigma_i}
\begin{aligned}
    \Delta P^{\rm err}_{k,\sigma_i}=m_{\sigma_i}\sum_{l\in{\rm cell}k \atop \sigma_l=\sigma_i}\Delta_kN_l\vec c_l-\Delta P_{k,\sigma_i},
\end{aligned}
\end{equation}
where $\Delta P_{k,\sigma_i}$ is computed by Eq.~\eqref{eq:momentumincrement-sigma_i} using $n_{{\rm tr},k,\sigma_i}$, $n_{{\rm tr},k,\sigma_j}$, $\vec u_{{\rm tr},k,\sigma_i}$, $\vec u_{{\rm tr},k,\sigma_j}$ in place of $n_{\sigma_i}$, $n_{\sigma_j}$, $\vec u_{\sigma_i}$, $\vec u_{\sigma_j}$, respectively. The error between the numerical energy increment and theoretical energy increment $\Delta E_{k,\sigma_i}$ of component $\sigma_i$ in the cell $k$ due to intermolecular collisions with all components is 
\begin{equation}\label{eq:energyincrementerr-sigma_i}
\begin{aligned}
    \Delta E^{\rm err}_{k,\sigma_i}=\dfrac{m_{\sigma_i}}{2}\sum_{l\in{\rm cell}k \atop \sigma_l=\sigma_i}\Delta_kN_l(\vec c_l)^2-\Delta E_{k,\sigma_i},
\end{aligned}
\end{equation}
where $\Delta E_{k,\sigma_i}$ is computed by Eq.~\eqref{eq:energyincrement-sigma_i} using $n_{{\rm tr},k,\sigma_i}$, $n_{{\rm tr},k,\sigma_j}$, $\vec u_{{\rm tr},k,\sigma_i}$, $\vec u_{{\rm tr},k,\sigma_j}$, $3k_{\rm B}T_{{\rm tr},k,\sigma_i}/2$, $3k_{\rm B}T_{{\rm tr},k,\sigma_j}/2$ in place of $n_{\sigma_i}$, $n_{\sigma_j}$, $\vec u_{\sigma_i}$, $\vec u_{\sigma_j}$, $\epsilon_{\sigma_i}$, $\epsilon_{\sigma_j}$, respectively.  

The above numerical errors are used to update the cell's variables $n_{{\rm tr},k,\sigma_i}$, $\vec u_{{\rm tr},k,\sigma_i}$ and $T_{{\rm tr},k,\sigma_i}$ at the end of each $\Delta t$ based on an auto-regulation scheme \cite{Li2012arXiv}:
\begin{equation}\label{eq:auto}
    \begin{cases}
    n_{{\rm tr,}k,\sigma_i}^{\rm new}=\dfrac{n_{{\rm tr,}k,\sigma_i}\Delta V_k-\Delta M^{\rm err}_{k,\sigma_i}/m_{\sigma_i}}{\Delta V_k} \\
    \vec u_{{\rm tr,}k,\sigma_i}^{\rm new}=\dfrac{n_{{\rm tr,}k,\sigma_i}\Delta V_k\vec u_{{\rm tr,}k,\sigma_i}-\Delta P^{\rm err}_{k,\sigma_i}/m_{\sigma_i}}{n_{{\rm tr,}k,\sigma_i}^{\rm new}\Delta V_k} \\
    T_{{\rm tr,}k,\sigma_i}^{\rm new}=\dfrac{[n_{{\rm tr,}k,\sigma_i}\Delta V_k(\dfrac{3k_{\rm B}T_{{\rm tr,}k,\sigma_i}}{2}+\dfrac{m_{\sigma_i}\vec u_{{\rm tr,}k,\sigma_i}^2}{2})-\Delta E^{\rm err}_{k,\sigma_i}]-n_{{\rm tr,}k,\sigma_i}^{\rm new}\Delta V_k\dfrac{m_{\sigma_i}(\vec u_{{\rm tr,}k,\sigma_i}^{\rm new})^2}{2}}{n_{{\rm tr,}k,\sigma_i}^{\rm new}\Delta V_k\dfrac{3k_{\rm B}}{2}},
    \end{cases}
\end{equation}
where $n_{{\rm tr,}k,\sigma_i}^{\rm new}, \vec u_{{\rm tr,}k,\sigma_i}^{\rm new}, T_{{\rm tr,}k,\sigma_i}^{\rm new}$ are the new values of number density $n_{{\rm tr,}k,\sigma_i}$, flow velocity $\vec u_{{\rm tr,}k,\sigma_i}$ and temperature $T_{{\rm tr,}k,\sigma_i}$ of the component $\sigma_i$ at the cell $k$, respectively. Then, the cell's auxiliary variables $\vec u^{\rm eq}_{{\rm tr,}k,\sigma_i}$ and $T^{\rm eq}_{{\rm tr,}k,\sigma_i}$ can be updated by Eqs.~\eqref{eq:multiueq} and \eqref{eq:multiepsiloneq}, where the updated discrete variables $n_{{\rm tr,}k,\sigma_i}^{\rm new}, \vec u_{{\rm tr,}k,\sigma_i}^{\rm new}, T_{{\rm tr,}k,\sigma_i}^{\rm new}$ are used to replace $n_{\sigma_i}$, $\vec u_{\sigma_i}$, $T_{\sigma_i}$, respectively.

%


\begin{thebibliography}{23}
\expandafter\ifx\csname natexlab\endcsname\relax\def\natexlab#1{#1}\fi
\providecommand{\bibinfo}[2]{#2}
\ifx\xfnm\relax \def\xfnm[#1]{\unskip,\space#1}\fi

\bibitem{CL1971}
\bibinfo{author}{Carlo~Cercignani}, \bibinfo{author}{Maria~Lampis},
\newblock \bibinfo{title}{Kinetic Models for Gas-surface Interactions},
\newblock \bibinfo{journal}{Transport Theory and Statistical Physics} \bibinfo{volume}{1(2)} (\bibinfo{year}{1971}) \bibinfo{pages}{101-114}.

\bibitem{CLL1991}
\bibinfo{author}{R.G.~Lord},
\newblock \bibinfo{title}{Some Extensions to the Cercignani-Lampis Gas-surface Scattering Kernel},
\newblock \bibinfo{journal}{Physics of Fluids} \bibinfo{volume}{3(4)} (\bibinfo{year}{1991}) \bibinfo{pages}{706-710}.

\bibitem{Bird1994}
\bibinfo{author}{Graeme~A.~Bird},
\bibinfo{title}{Molecular Gas Dynamics and the Direct Simulation of Gas Flows}, \bibinfo{publisher}{Clarendon Press, Oxford}, (\bibinfo{year}{1994}).

\bibitem{Li2011RGD}
\bibinfo{author}{Jun~Li},
\newblock \bibinfo{title}{Direct Simulation Method Based on BGK Equation},
\newblock in: \bibinfo{booktitle}{27th International Symposium on Rarefied Gas Dynamics},
\bibinfo{publisher}{AIP}, (\bibinfo{year}{2011}), \bibinfo{pages}{283-288 (\textit{presented first in ESPCI, Paris, 2009})}.

\bibitem{Li2012arXiv}
\bibinfo{author}{Jun~Li},
\newblock \bibinfo{title}{Comparison between the DSMC and DSBGK Methods},
\newblock \bibinfo{journal}{\href{http://arxiv.org/abs/1207.1040}{arXiv:1207.1040 [physics.comp-ph]}},
(\bibinfo{year}{2012}).

\bibitem{Li2014arXiv}
\bibinfo{author}{Jun~Li},
\newblock \bibinfo{title}{Improved Diffuse Boundary Condition for the DSBGK Method to Eliminate the Unphysical Density Drift},
\newblock \bibinfo{journal}{\href{http://arxiv.org/abs/1403.3923}{arXiv:1403.3923 [physics.comp-ph]}},
(\bibinfo{year}{2014}).

\bibitem{LiSultan2015JNGSE}
\bibinfo{author}{Jun~Li}, \bibinfo{author}{Abdullah~S.~Sultan},
\newblock \bibinfo{title}{Klinkenberg Slippage Effect in the Permeability Computations of Shale Gas by the Pore-scale Simulations},
\newblock \bibinfo{journal}{Journal of Natural Gas Science and Engineering,} (\bibinfo{year}{2016}), \bibinfo{pages}{in press}.

\bibitem{LiSultan2015IPTC}
\bibinfo{author}{Jun~Li}, \bibinfo{author}{Abdullah~S.~Sultan},
\newblock \bibinfo{title}{Permeability Computations of Shale Gas by the Pore-scale Monte Carlo Molecular Simulations},
\newblock in: \bibinfo{booktitle}{International Petroleum Technology Conference},
(\bibinfo{year}{2015}), \bibinfo{pages}{IPTC-18263-MS}.

\bibitem{LiCai2017arXiv}
\bibinfo{author}{Jun~Li}, \bibinfo{author}{Chunpei~Cai},
\newblock \bibinfo{title}{Numerical Study on Thermal Transpiration Flows Through a Rectangular Channel},
\newblock \bibinfo{journal}{\href{https://arxiv.org/abs/1708.08105}{arXiv:1708.08105 [physics.flu-dyn]}},
(\bibinfo{year}{2017}).

\bibitem{BGK1954}
\bibinfo{author}{P. L.~Bhatnagar}, \bibinfo{author}{E. P.~Gross}, \bibinfo{author}{M.~Krook},
\newblock \bibinfo{title}{A Model for Collision Processes in Gases. I. Small Amplitude Processes in Charged and Neutral One-Component Systems},
\newblock \bibinfo{journal}{Physical Review} \bibinfo{volume}{94(3)} (\bibinfo{year}{1954}) \bibinfo{pages}{511-525}.

\bibitem{VincentiKruger1965}
\bibinfo{author}{Walter~G.~Vincenti}, \bibinfo{author}{Charles~H.~Kruger, Jr.},
\bibinfo{title}{Introduction to Physical Gas Dynamics}, \bibinfo{publisher}{John Wiley \& Sons}, (\bibinfo{year}{1965}).

\bibitem{Shen2005}
\bibinfo{author}{Ching~Shen},
\bibinfo{title}{Rarefied Gas Dynamics: Fundamentals, Simulations and Micro Flows}, \bibinfo{publisher}{Springer}, (\bibinfo{year}{2005}).

\bibitem{AndriesAokiPerthame2002}
\bibinfo{author}{Pierre~Andries}, \bibinfo{author}{Kazuo~Aoki}, \bibinfo{author}{Benoit~Perthame},
\newblock \bibinfo{title}{A Consistent BGK-Type Model for Gas Mixtures},
\newblock \bibinfo{journal}{Journal of Statistical Physics} \bibinfo{volume}{106} (\bibinfo{year}{2002}) \bibinfo{pages}{993-1018}.

\end{thebibliography}
\end{document}